\documentclass[baaa]{baaa}

\usepackage[pdftex]{hyperref}
\usepackage{subfigure}
\usepackage{natbib}
\usepackage{helvet,soul}
\usepackage[font=small]{caption}

\contriblanguage{0}

\contribtype{8}

\thematicarea{99}

\title{Magnetic Fields in Cosmic Voids}

\titlerunning{Magnetic Fields in Voids}

\author{
A. M. Rodr\'iguez-Medrano\inst{1},
F. A. Stasyszyn \inst{1,2},
D. J. Paz \inst{1,2}
\&
V. Springel \inst{3}
}

\authorrunning{Rodr\'iguez-Medrano et al.}

\contact{arodriguez@unc.edu.ar}

\institute{
Instituto de Astronom\'ia Te\'orica y Experimental, CONICET--UNC, Argentina
\and   
Observatorio Astron\'omico de C\'ordoba, UNC, Argentina
\and
Max-Planck-Institut für Astrophysik, Karl-Schwarzschild-Straße 1, 85741 Garching, Germany
}

\abstract{Magnetic fields are one of most concealed components of the universe. They are observed as part of the intergalactic medium and on galaxy cluster scales, however their origin and evolution is unclear. In this work we use the IllustrisTNG simulations to investigate the effects of magnetic fields in cosmic voids, the least dense regions of the universe. We find that, under the hypothesis of a uniform primordial magnetic field, the voids still reflect the primordial properties of the fields. On the other hand, the galaxies in their interior acquire weaker magnetic fields than galaxies in denser environments.}

\resumen{Los campos magn\'eticos constituyen una de las componentes más ocultas del universo. Estos son observados tanto como parte del medio intergal\'actico hasta en escalas de c\'umulos de galaxias, sin embargo su origen y evolución no est\'a claro. En este trabajo utilizamos las simulaciones IllustrisTNG para investigar los efectos de estos campos magn\'eticos en los vac\'ios c\'osmicos, las regiones m\'as subdensas del universo. Encontramos que, bajo la hip\'otesis de un campo magn\'etico primordial, los vac\'ios conservan propiedades primordiales de los campos. Por otro lado, galaxias en sus interiores conservan campos magn\'eticos m\'as tenues que galaxias en entornos m\'as densos. 
}


\begin{document}

\maketitle

\section{Introduction}\label{S_intro}

In recent years,  it has become possible to detect magnetic fields with various techniques and at different scales in the universe \citep{Vallee1998,Carilli2002, Beck_book2013}. However, the origin of these fields is a source of constant debate. Under the hypothesis of a cosmological origin of the magnetic fields, the latest generation of cosmological simulations has introduced a weak primordial magnetic field in the initial conditions and showed that the non-linear processes that occur during the evolution of structures are responsible for amplifying the magnetic field up to the currently observed values \citep{Marinacci2015, Marinacci2018}.

If the primordial origin of the fields is true, regions of low matter density are natural candidates to harbour some primordial remnants of these fields. This is because, in low-density environments, the field would not have been amplified or modified by halo collapse and astrophysical processes, but would rather for the most part evolve adiabatically \citep{Marinacci2015}. In this work, we study the magnetic fields in cosmic voids, the most underdense regions of the universe, in order to constrain the imprints of the primordial magnetic field in these environments and in the galaxies within these regions.


\section{Methods}

In this work we analyse the TNG100 simulation of the IllustrisTNG project \citep{TNG_datarelease_dylan2018,Dylan2018, Marinacci2018, Naiman2018, Pillepich2018, Springel2018}. This simulation follows a volume $\sim 110 \,\rm Mpc $ on a side and simulates both $1820^{3}$ dark matter particles and gas, thus achieving a mass resolution of $M_{\rm DM}\sim 7\times10^{6} M_{ \odot}$ and $M_{\rm gas}\sim 1\times10^{6}M_{\odot}$ for dark matter and gas, respectively. The simulation is magneto-hydrodynamic and includes state of the art subgrid models of galaxy formation physics. In particular, magnetic fields evolve from a primordial seed strength of $10^{-14}$ (comoving) Gauss placed uniformly in the $z$-direction in the initial condition at reshift $z=127$. This implementation of a homogeneous field does not have a solid physical motivation, but it guarantees the necessary condition of having an initial B-field free of divergences. Details about this implementation can be found in \citet{Marinacci2015, Marinacci2018}.

Regarding the voids, they were identified using the Popcorn code \citep{Paz2022} that use the void definition due by \citet{Padilla2005}.
In this way, voids are defined as spheres in the galaxy distribution with an integrated density contrast ($\Delta$) below a certain limit (see the referenced articles for more details). We identified voids for the sample of subhaloes with masses $M>10^{10}M_{\odot}$ \citep[the halos and subhalos were identified with {\small SUBFIND}][]{Springel2001}. The catalogue consists of spherical voids of radii $\sim r_{\rm void}>3$ Mpc identified with an integrated density contrast $\Delta<-0.9$.

\section{Results}\label{sec:results}
 
The implementation of the magnetic field in the initial conditions as a field aligned with an axis imprints directional biases when one looks at it in regions of low-density \citep{Marinacci2015}. Naturally, voids are ideal regions to test for this bias. For this purpose we calculate the magnetic field profile in each direction $(x, y, z)$. To calculate the profile, we compute the net field in each direction (i.e.~$B_x, B_y, B_z$)  in shells of gas cells centred around the middle of the void. In each shell, we compute the average value $B_i^2$ of each Cartesian component. 

\begin{figure}
    \centering
    \includegraphics[width=\columnwidth]{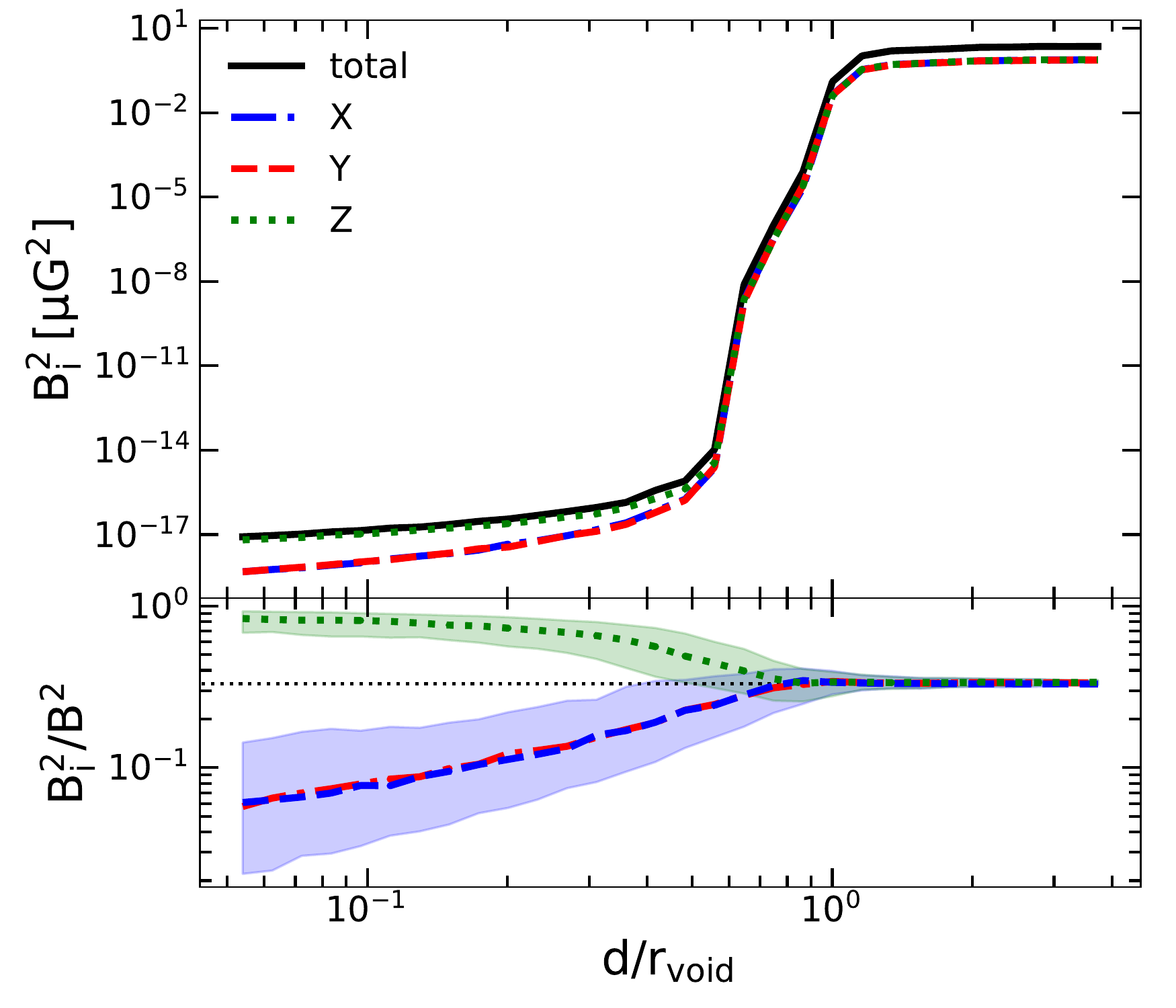}
    \caption{
    The figure shows in the \emph{upper panel} the $B_i^2$ value as a function of the distance to the void center, where $i$ is the $x$, $y$ or $z$ component of the magnetic field vector.
    With red, blue and green colours we show the median value of the $x$, $y$ and $z$ components, respectively. 
    In the \emph{bottom panel}, the figure shows the median $B_i^{2}/\left<B^{2}\right>$ value as a function of the distance to the void center.
    The shaded areas indicate 25 and 75 percentiles. The black dotted line shows the 0.33 value. }
    \label{B2perf}
\end{figure}

In Fig.~\ref{B2perf} we show in the upper panel the $\left<B^{2}\right>$ profile for voids (continuous  black line) and for each component $B_x$ (dot-dashed-blue), $B_y$ (dashed-red), and $B_z$ (dotted-green). Each profile was calculated considering the median value of all profiles. As it can see in the figure, the inner regions of voids (for distances $d<r_{\rm void}$) have a deficit of magnetic field relative to the outside void regions. In the inner regions of the voids, the total value of the magnetic field is similar to the magnetic field in the $z$ direction. At a distance close to the void radius $r_{\rm void}$, the dominance of this component disappears. 

This can be seen most clearly in the bottom panel of the figure. In this panel we divide each component of the magnetic field by the total field. Towards values of $\sim d<0.8$, the figure indicates differences in the median of the magnetic field values, resulting in an alignment of the magnetic field in the $z$-direction (the primordial direction). Towards values beyond this limit, the alignment is not perceptible and all values converge to 0.33, indicating that the direction is random.

Within the halos, the magnetic fields are amplified by several orders of magnitude by non-linear processes and dynamo effects, and the primordial direction of the initial conditions is lost \citep{Marinacci2015}. However, it is interesting to consider if there is any dependence of the magnitude of the magnetic field on the environment in which the galaxies reside.  For the galaxies identified in the simulation we have two measurements of the magnetic field, the $B_{\rm disk}$ and $B_{\rm halo}$ values. Each is defined as the volume-weighted value of $|B|$ for the gas cell in the halo ($B_{\rm halo}$) and within twice the radius of the  stellar half mass radius ($B_{\rm disk}$). Thus, the first represents the field associated with the halo and the second that associated with the galaxy.

\begin{figure*}
    \centering
    \includegraphics[width=\linewidth]{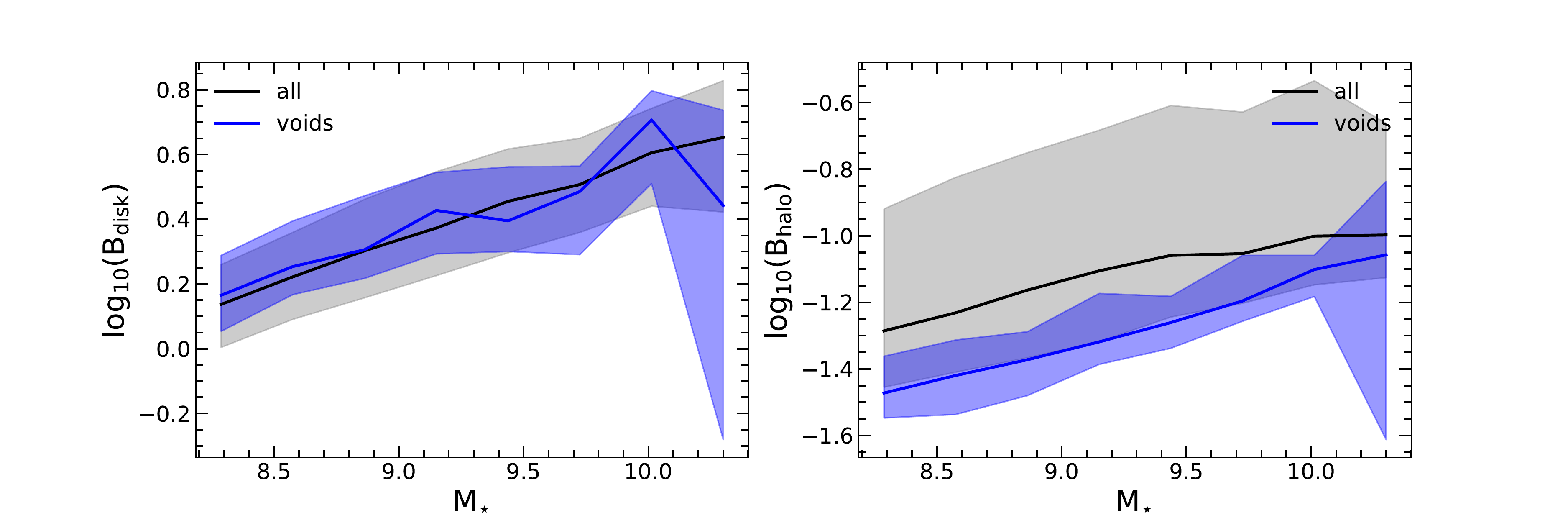}
    \caption{In the \emph{left panel} we show the $\log_{10}(B_{\rm disk})$ as a function of the stellar mass, and in the \emph{right panel} the $\log_{10}(B_{\rm halo})$ also as a function of the stellar mass. With blue colours we show the median value for galaxies within voids, and with black for the full sample of galaxies. The shaded areas indicate 25 and 75 percentiles. }
    \label{B-mst}
\end{figure*}

In Fig.~\ref{B-mst} we show the median $B_{\rm disk}$ and $B_{\rm halo}$ values as a function of the stellar mass of the galaxy $(M_{\star})$ (left and right panels, respectively). The blue line shows the behaviour of galaxies within the voids (at a distance $d<r_{\rm void}$), and the complete sample of galaxies is shown in black. When  $B_{\rm disk}$ is analysed, it is evident that the greater the mass of the galaxy, the greater the magnetic field. However, there are no differences between galaxies inside or outside the voids. In the case of $B_{\rm halo}$, we generally find smaller values than for $B_{\rm disk}$, which is easily understood because the highest values of the magnetic field are expected in the centre of the halos, where the density is highest. By comparing the value of $B_{\rm halo}$ as a function of the mass of the halos for halos in voids and in the general sample, we find that those in voids have a lower magnetic field value. Although the corresponding differences are small, they can be seen across the entire range of masses that we analysed.

\section{Conclusions}\label{sec:conclusions}

In this brief article, we have studied voids as potential regions to harbor imprints of the primordial initial conditions of the magnetic fields in the universe. Through the profiles of the magnetic field in cosmic voids, we observe how in their internal regions the field maintains its original seed direction (as imposed in the initial conditions). Although this direction is artificial due to the seed model of the magnetic field, the fact that it has only evolved by cosmic expansion suggests that if fields are detected there, they could still hold information and reflect properties about the primordial seed field. 

On the other hand, we find that, at a given stellar mass, galaxies that inhabit voids have a lower magnetic field strength compared to galaxies in large-scale denser environments. These differences are only appreciable when the magnetic field of the entire halo is considered. When the fields of the inner regions of the halo are considered (where the galaxies are located), the differences are lost. This indicates that the non-linear processes and dynamos that are efficient in amplifying magnetic fields during galaxy formation also erase environmental differences in magnetic field signatures.

\begin{acknowledgement}
\texttt{This project has received financial support from
the European Union’s Horizon 2020 Research and Innovation programme under the Marie Sklodowska-Curie grant agreement number 734374 - project acronym: LACEGAL.}

\end{acknowledgement}


\bibliographystyle{baaa}
\small
\bibliography{example}
 
\end{document}